\documentclass[journal]{IEEEtran}

\usepackage{lineno,hyperref}
\usepackage{caption}
\usepackage{graphicx, subcaption}         
\usepackage{amsfonts}
\usepackage{amssymb}
\usepackage{amsmath}
\usepackage{booktabs}
\usepackage{bm}
\usepackage{algorithm}
\usepackage{savesym}
\savesymbol{AND}
\usepackage{algpseudocode}
\usepackage{color} 
\usepackage{enumitem}

\newtheorem{problem}{Problem}

\newcommand{\Rnum}[1]{\uppercase\expandafter{\romannumeral #1\relax}}
\newcommand{\rnum}[1]{\lowercase\expandafter{\romannumeral #1\relax}}
  
\DeclareMathOperator{\atantwo}{atan2}

\usepackage[numbers,sort&compress]{natbib}

\ifCLASSINFOpdf
\else
\fi
\begin{document}
\title{Economic MPC-based planning for marine vehicles: Tuning safety and energy efficiency}

\author{
	\vskip 1em
	
	Haojiao~Liang, 
	Huiping~Li, Jian~Gao, Rongxin~Cui, and Demin~Xu

	\thanks{
		Haojiao~Liang, Huiping~Li, Jian Gao, Rongxin~Cui, and Demin~Xu are with the School of Marine Science and Technology, Northwestern Polytechnical University, Xi'an, 710072, China, (corresponding author: Huiping Li, phone: 86-029-88492611; fax: 86-029-88495278; e-mail: lihuiping@nwpu.edu.cn)
	}
}

\maketitle
	
\begin{abstract}
Energy efficiency and safety are two critical objectives for marine vehicles operating in environments with obstacles, and they generally conflict with each other.  
In this paper, we propose a novel online motion planning method of marine vehicles which can make trade-offs between the two design objectives based on the framework of economic model predictive control (EMPC).
Firstly, the feasible trajectory with the most safety margin is designed and utilized as tracking reference. Secondly, the EMPC-based receding horizon motion planning algorithm is designed, in which the practical consumed energy and safety measure (i.e., the distance between the planning trajectory and the reference) are considered. Experimental results verify the effectiveness and feasibility of the proposed method.
\end{abstract}



\definecolor{limegreen}{rgb}{0.2, 0.8, 0.2}
\definecolor{forestgreen}{rgb}{0.13, 0.55, 0.13}
\definecolor{greenhtml}{rgb}{0.0, 0.5, 0.0}

\section{Introduction}
Marine vehicles find wide applications in civilian and military tasks, such as environment monitoring, resource exploration, search, and rescue. 
It is well recognized that motion planning is one of the most important problems for marine vehicles, which are critical to promote their intelligence and reliability. Essentially speaking, motion planner (i.e., the physical module for motion planning) is to find state and input pairs enabling marine vehicles to move from one location to another while avoiding obstacles. In this way, motion planners provide references to the controllers of marine vehicles, and instruct them to execute various tasks safely, autonomously, and efficiently.

In the literature, there are three conventional and popular approaches to the motion planning problem, namely, the geometric model-based method~\cite{niu2018energy,chi2021generalized,wu2021long}, the potential field method~\cite{khatib1986real,huang2019motion,malone2017hybrid}, and the sampling-based method~\cite{lavalle1998rapidly,karaman2011sampling,hu2020efficient}. The geometric model-based method mainly uses roadmap and cell decomposition techniques to represent paths, but it cannot address the dynamics of marine vehicles. The potential filed method is easily trapped in local minima. Though the sampling-based method can overcome the drawbacks of those two approaches, it might not converge to the optimal points with finite number of samples. In addition, it might not be efficient for solving planning problems with multi-objectives under constrained conditions.

To deal with complex constraints and design objectives, the metaheuristic-based planning algorithm has been developed. 
For example, the genetic algorithm was utilized to solve the path planning problem for marine vehicles under current environments in~\cite{alvarez2004evolutionary} and~\cite{niu2020energy}, where the economic indexes were designed by exploiting the hydrodynamic drag. The particle swarm optimization method was introduced in~\cite{ma2018multi} to obtain an optimal path by taking the influence of ocean current, path length, simple kinematic model, and traveling time into consideration. However, these methods might be computational expensive for real marine vehicles with limited computation power.

The MPC-based approach is potentially promising in planning problems because of its receding horizon mechanism, which naturally adapts to the limitation of sensor range and online implementation.
For example, a motion planning method implemented in the receding horizon fashion was designed by resorting to the MPC technique in~\cite{kufoalor2020autonomous}, where the objective function consisted of penalties for colliding with obstacles and violating the established rules.  
The robust MPC method was also utilized to determine the optimal motion under the known current environment in~\cite{huynh2015predictive}, where the objective was to minimize the energy consumption associated with drag forces, and the optimization problem with predetermined input sets and kinematic constraints was solved by the A* like method.

Although great progress has been made in the MPC-based motion planning problem, there is lacking an online planning framework simultaneously addressing the power consumption and safety margin for marine vehicles.
In this paper, we investigate the online motion planning problem of a marine vehicle in cluttered environments, where the marine vehicle is assigned to visit predefined waypoints closing to the interesting areas while avoiding obstacles with safety margin, satisfying physical constraints, and saving energy as much as possible. To that end, we develop an EMPC-based motion planning method for marine vehicles and provide explicit implementation algorithms for real time applications. 

The contributions of the paper are as follows:

\begin{itemize}
	\item 
	A novel method for design smooth trajectories with maximum safety margin is proposed by taking advantage of Voronoi diagrams and B{\'e}zier curves.
	In particular, the multiscale priority criterion is firstly designed to search the paths with maximum safety margin. The new smoothing method for the environments with both dense and sparse obstacles is proposed to smooth the designed paths. The designed path is then converged into the smooth trajectory with safety margin for the tracking reference.  
	
    \item An EMPC-based online motion planning framework for marine vehicles that can enable the balance between the power consumption and safety margin is developed. In this framework, the actual power consumption of marine vehicles is modeled as an economic cost term in the objective function of EMPC, and the distance between the planned trajectory and the reference is characterized as a term of safety measure, and the obstacle avoidance and fulfillment of system constraints are explicitly considered in the optimization problem. In this way, the planned motions are collision-free, energy-efficient, feasible, and easy for tracking.
	
\end{itemize}


Notation:
For a vector $v$, $\|v\|$ denotes its Euclidean norm. For a real number $v$, $|v|$ stands for its absolute value. For a positive real number $v$, $v^{1.5}$ represents $\sqrt{v^3}$. The symbols $G^0$ and $G^1$ denote the curves touching, and sharing common tangent directions at join points, respectively. For the points $p_0$, $p_1$, and $p_2$, $\overline{p_0 p_1}$, $\overrightarrow{p_0 p_1}$, and $\angle{p_0 p_1 p_2}$ denote a straight line, a directed line, and the included angle of $\overrightarrow{p_1 p_0}$ and $\overrightarrow{p_1 p_2}$, respectively.

\section{Problem formulation}\label{sec_pro_for}
\begin{figure}[htb] 
	\centerline{\includegraphics[width=2.25in]{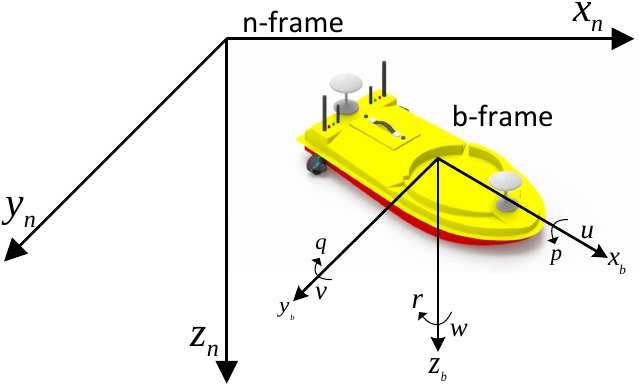}}   
	\caption{The vehicle in the $b$- and $n$-frames: ($u$,$v$,$w$) and ($p$,$q$,$r$) are the vehicle's linear and angular velocities, respectively.} 
	\label{fig-model} 
\end{figure}

The marine vehicle is shown in Fig.~\ref{fig-model}, where the $n$ (north-east-down) frame $x_n y_n z_n$ defined as the tangent plane on the surface of the earth with $x$, $y$, and $z$ axes pointed towards north, east, and downwards, respectively, and the $b$ (body-fixed) frame $x_b y_b z_b$ is moving with the vehicle, with $x$, $y$, and $z$ axes directed from aft to fore, from port-side to starboard, and from top to bottom, respectively.
The three degrees of freedom horizontal motion, containing motion in surge, sway and yaw, is considered here, for it plays the leading role for saving energy in motion planning tasks.

For simplicity, it is assumed that the marine vehicle is well symmetric and moves at low speed. As a result, the off-diagonal elements of inertial and damping matrices and nonlinear terms of the damping matrix are neglected. Then, the dynamic equations are described as:
\begin{equation}\small \label{eq_dynamic1}
	\begin{aligned}
		\begin{pmatrix} \dot{x}\\ \dot{y}\\ \dot{\psi} \end{pmatrix}=
		\begin{pmatrix} \cos{\psi}&-\sin{\psi}&0\\ \sin{\psi}&\cos{\psi}&0\\ 0&0&1 \end{pmatrix}
		\begin{pmatrix} u\\ v\\ r \end{pmatrix},
	\end{aligned}
\end{equation}
\begin{equation} \small \label{eq_dynamic2}
	\begin{cases}
		M_{1}\dot{u}=&M_{2}vr - D_{1}u + X  \\
		M_{2}\dot{v}=&- M_{1}ur - D_{2}v    \\
		M_{3}\dot{r}=&(M_{1}-M_{2})uv - D_{3}r + N,
	\end{cases}
\end{equation}
where $M_{i}$ and $D_{i}$ ($i=1,\cdots 3$) are diagonal elements of the system inertial matrix and linear parts of the damping matrix, respectively; vectors $(x, y, \psi)$ and $(u, v, r)$ are vehicle's posture in the $n$-frame and its velocity in the $b$-frame, respectively; $X$ and $N$ denote the force and torque provided by actuators, respectively. For more parameters description, please see~\cite{fossen2011handbook}. 
For the sake of clarity, the model is written as $\dot{\bm{x}}=f(\bm{x},\bm{\tau})$, and the vehicle's position is denoted as $\bm{x}_p$, where $\bm{x} \triangleq [x, y, \psi, u, v, r]^{T}$ and $\bm{\tau} \triangleq [X, N]^{T}$.

To constrain the jerk of motion, an extended system is constructed by taking the vector $\bm{\tau}$ as an augmented state and setting $\bm{\tau}_{\delta}\triangleq[X_{\delta},N_{\delta}]^T$ (i.e., the rate of change of $\bm{\tau}$) as control input. As a result, the extended system is described as:
\begin{equation} \small \label{eq_extsystem}
	\dot{\bm{x}}_{\text{aug}} \triangleq
	\begin{bmatrix} \dot{\bm{x}} \\ \dot{\bm{\tau}} \end{bmatrix} =
	\begin{bmatrix}  f(\bm{x}, \bm{\tau}) \\ \bm{\tau}_{\delta} \end{bmatrix} \triangleq
	f_{\text{aug}}(\bm{x}_{\text{aug}}, \bm{\tau}_{\delta}),
\end{equation} 
where the new variables $\bm{x}_{\text{aug}}$ and $\bm{\tau}_{\delta}$ are the state and control input of the augmented system \eqref{eq_extsystem}, respectively.

The motion planning problem is to design an EMPC-based online planning framework from a starting point to a destination while 1) satisfying system constraints, 2) avoiding collisions with environmental obstacles, and 3) having the capability to tune safety margin and energy efficiency.



\section{Trajectory design of safety margin}\label{sec_trajectory}

\subsection{Search the most safe path}
\begin{figure}[htb]
	\centerline{\includegraphics[width=2.5in, height=1.25in]{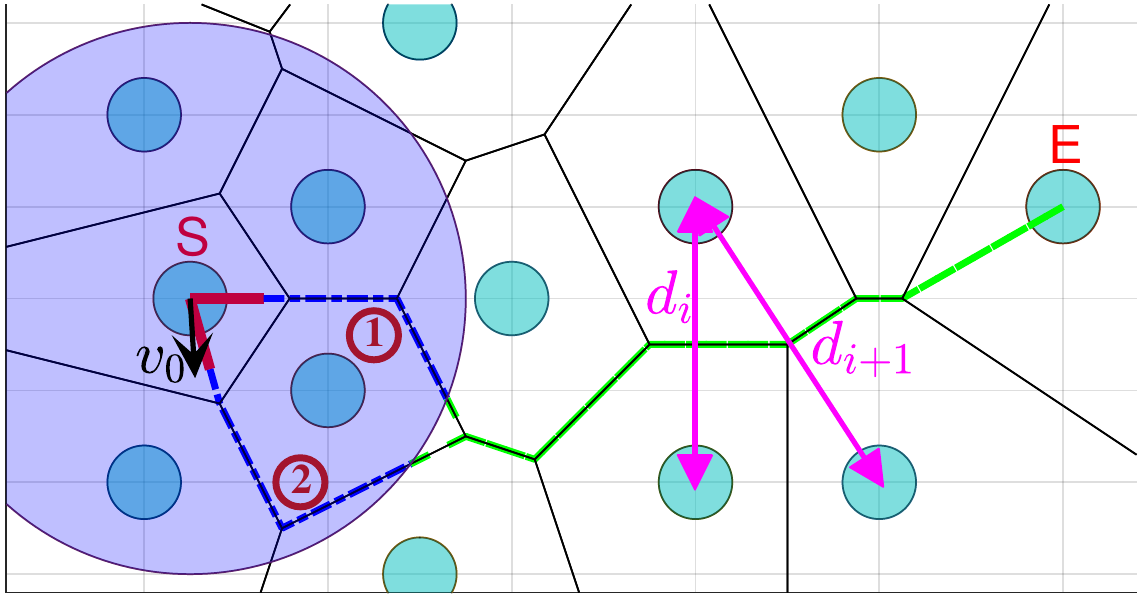}}   
	\caption{Path selection: The first part to be traversed (red line), the middle part within the range of detection (blue line), and the remaining unknown part (green line) are used to measure the cost of jerk, acceleration, and length, respectively.} 
	\label{fig-sel-direction} 
\end{figure}

To obtain the path with maximum safety margin, the geometrical environment is described by the Voronoi diagram as shown in Fig.~\ref{fig-sel-direction}, where turquoise points and black lines are Voronoi sites and edges, respectively. 
As shown in Fig.~\ref{fig-sel-direction}, the shortest path is labeled `1', and it is obtained by removing edges with narrow passages judging by the distance $d_i$, and searching with the Dijkstra's algorithm, where the black arrow line denotes the direction of velocity.


From Fig.~\ref{fig-sel-direction}, we see that there is a big jerk in the initial part of the searched path of label 1.
To obtain a more suitable path, a multiscale priority selection method is designed to evaluate the paths, and the optimal one is label 2. The details are as follows: 1) Search the shortest paths through vertices of the site `S'; 2) Evaluate the searched paths according to the multiscale priority criterion:
\begin{equation} \small \label{eq_j_k}
	\mathcal{J}_{\rho_1}-\mathcal{J}_{\rho_2} \triangleq  J_{\Delta} + w_1 \Pi(J_{\Delta})A_{\Delta} + w_2 \Pi(J_{\Delta}) \Pi(A_{\Delta})L_{\Delta},
\end{equation}
where $\mathcal{J}_{(\cdot)}$ stands for the total cost of a path (e.g., paths $\rho_1$ and $\rho_2$); $w_{1}$ and $w_{2}$ are two constants; 
$J_{\Delta} \triangleq J_{\rho_1}-J_{\rho_2}$, $A_{\Delta} \triangleq A_{\rho_1}-A_{\rho_2}$ and $L_{\Delta} \triangleq L_{\rho_1}-L_{\rho_2}$, where $J_{(\cdot)}$, $A_{(\cdot)}$ and $L_{(\cdot)}$ denote jerk of the first part, acceleration of the middle part and length of the remaining part, respectively; the function $\Pi$ is defined as:
\begin{equation} \small 
	\Pi(x) = \begin{cases}
		1, &x\in[0,c), \\
		0, &x\in(-\infty,-c]\cup[c,\infty), \\
		-1,&(-c,0), \\
	\end{cases}
\end{equation}
where the constant $c$ is the threshold of $\Pi$;
the calculations of $J_{(\cdot)}$ and $A_{(\cdot)}$ are described as follows.

(1) \textit{Calculation of jerk}:
The value of the first term in~\eqref{eq_j_k} (i.e., $J_{(\cdot)}$ defined as $\sum_{k=1}^{2}{\int_{0}^{T_k}{ j_k^2(t)/T_k dt}}$, where $j_{(\cdot)}$ denotes the cost along axes and $T_k$ is the during time), is denoted as $J_{(\cdot)}=\sum_{k=1}^{2} \gamma_k^2+\beta_k \gamma_k T_k+\frac{1}{3}\beta_k^2 T_k^2 + \frac{1}{3}\alpha_k \gamma_k T_k^2 + \frac{1}{4}\alpha_k \beta_k T_k^3 + \frac{1}{20}\alpha_k^2 T_k^4$, where $\alpha_k$, $\beta_k$ and $\gamma_k$ are determined as~\cite{mueller2015computationally} 
\begin{equation}  \small \label{eq_three_constant}
	\begin{bmatrix} \alpha_k \\ \beta_k \\ \gamma_k \end{bmatrix} = 
	\frac{1}{T_k^5} \begin{bmatrix} 320 & -120T_k \\ -200T_k & 72T_k^2 \\ 40T_k^2 & -12T_k^3 \end{bmatrix}
	\begin{bmatrix} \Delta p_k \\ \Delta v_k   \end{bmatrix},
\end{equation}
where $\Delta v_k \triangleq v_k^f-(a_k^0 T_k+v_k^0)$ and $\Delta p_k \triangleq p_k^f-(\frac{1}{2}a_k^0 T_k^2+v_k^0 T_k+p_k^0)$; $p_k$, $v_k$ and $a_k$ are initial states of position, velocity, and acceleration, respectively. Their initial and terminal states are marked with the superscripts 0 and $f$, respectively; $k=1$ and $k=2$ denote the costs along axes $x$ and $y$, respectively. 

(2) \textit{Calculation of acceleration}: 
To calculate the second term in~\eqref{eq_j_k} (i.e., $A_{(\cdot)}$ defined as $\sum_{k=1}^{2}\int_{0}^{T_k}a_k^2(t)/T_k dt$), we first define the state $s_k\triangleq[p_k,v_k]^T$ and its derivative $\dot{s}_k=[v_k,a_k]^T$. After further defining the costate $\lambda\triangleq[\lambda_1,\lambda_2]^T$, a Hamiltonian function can be constructed as $H(T_k, \lambda, \dot{s}_k) \triangleq \frac{1}{T_k}a_k^2 + \lambda^T \dot{s}_k$. According to the Pontryagin's minimum principle, trajectories of the costate, the optimal input and state can be described as: 
\begin{equation*} \label{eq_lmd_ast2} \small
	\lambda(t) = \frac{1}{T_k} \begin{bmatrix} -2\alpha_k \\ 2\alpha_k t+2\beta_k\end{bmatrix},
	\quad
	a_k^{\ast}(t) = -\lambda_2(t)T_k/2,
\end{equation*}
and
\begin{equation} \small \label{eq_s_ast2} 
	s_k^{\ast} = 
	\begin{pmatrix} -\alpha_k t^3/6-\beta_k t^2/2+v_k^0 t + p_k^0 \\ -\alpha_k t^2/2-\beta_k t+v_k^0\end{pmatrix},
\end{equation}
where~\eqref{eq_s_ast2} is obtained by the integration of $a_k^{\ast}(t)$; $\alpha_k$ and $\beta_k$ can be obtained by substituting $p_k^f$ and $v_k^f$ into \eqref{eq_s_ast2}. After rearranging, we get
\begin{equation} \small \label{eq_three_constant2} 
	\begin{bmatrix} \alpha_k \\ \beta_k \end{bmatrix} = 
	\frac{1}{T_k^3} \begin{bmatrix} 12 & -6T_k \\ -6T_k & 2T_k^2 \end{bmatrix}
	\begin{bmatrix} \Delta p_k \\ \Delta v_k   \end{bmatrix},
\end{equation}
where $\Delta p_k\triangleq p_k^f-(v_k^0 T_k+p_k^0)$ and $\Delta v_k\triangleq v_k^f-v_k^0$. After substituting the value of $\alpha_{k}$ and $\beta_{k}$ into~\eqref{eq_j_k}, we get $A_{(\cdot)}= \sum_{k=1}^{2}\frac{1}{3}\alpha_k^2 T_k^2 + \alpha_k \beta_k T_k + \beta_k^2$.

\subsection{Smooth the obtained path} \label{sec_smooth}
It is necessary to smooth the obtained path because there are many unnecessary jerks that will lead to time and energy wasting.
Considering prominent advantages of the B{\'e}zier curves, it is applied to smooth the path. Inspired by~\cite{choi2010real}, a new smoothing method for the environment with both dense and sparse obstacles is proposed. The smoothing procedure is carried out by two steps: Smooth the first B{\'e}zier curve and then the remaining ones.


\textbf{Step 1: Smooth the first B{\'e}zier curve}

We firstly smooth the first B{\'e}zier curve, in which we consider two situations in terms of obstacle densities.

\textbf{1) Design the first B{\'e}zier curve for the environment with dense obstacles}:

 For clarity, all the design situations for the first curve are divided into four cases according to the relative position of the line $\overline{p_0 p_1}$ and the points $\tilde{q}_1$ and $p_2$, where $p_{0\thicksim 2}$ are the first three waypoints, and $\tilde{q}_1$ is a point on the extension line along the direction of the initial velocity $\bm{v}_0$, where $\bm{v}_0\triangleq[u_0,v_0]^T$.

\begin{figure}[htb]
	\centering
	\begin{minipage}[t]{0.155\textwidth}
		\centerline{\includegraphics[height=4cm]{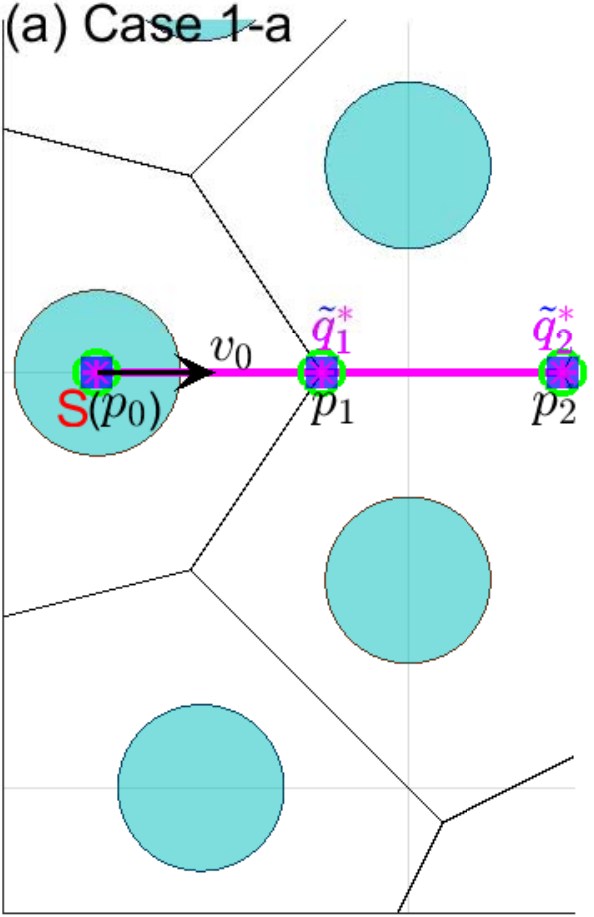}}   
	\end{minipage}
	\begin{minipage}[t]{0.155\textwidth}
		\centerline{\includegraphics[height=4cm]{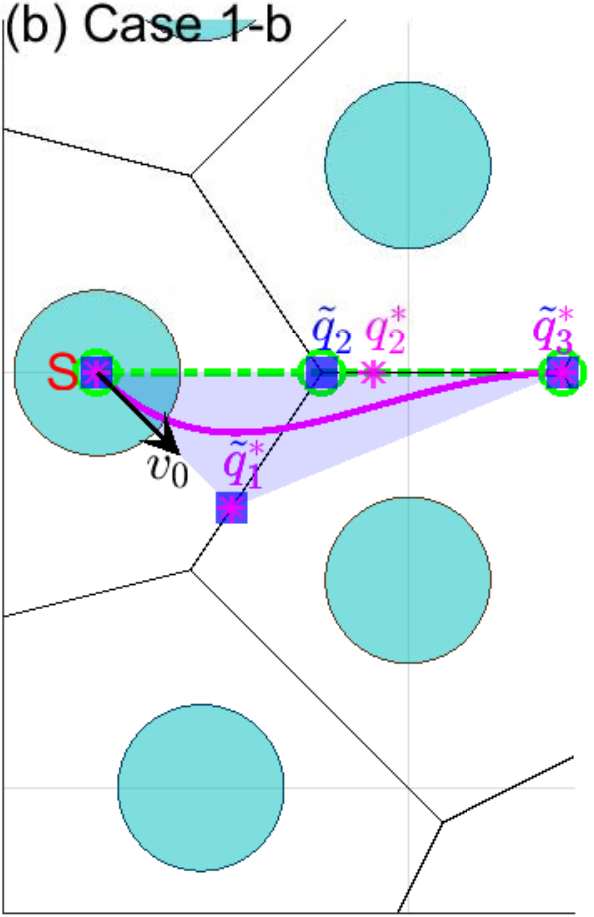}}   
	\end{minipage}
	\begin{minipage}[t]{0.155\textwidth}
		\centerline{\includegraphics[height=4cm]{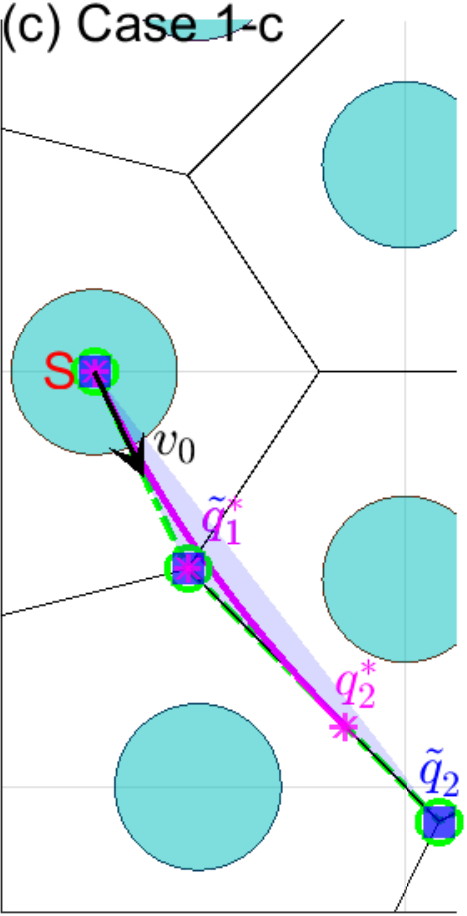}}   
	\end{minipage}
	\caption{Case 1: The three subgraphs correspond to Case1-a, Case 1-b and Case 1-c, respectively. The points marked with green circle, blue square and magenta asterisk are control points $p_{(\cdot)}$, candidate control points $\tilde{q}_{(\cdot)}$ and the optimal control points $q_{(\cdot)}^{\ast}$, respectively. 
		Shadow areas are the convex hulls constructed by $p_0$ and $\tilde{q}_{(\cdot)}$.} 
	\label{fig-Case-1} 
\end{figure}

\hangafter=1 \hangindent=3.5em \noindent
Case 1: The point $\tilde{q}_1$ and/or $p_2$ is on $\overline{p_0 p_1}$, and the angle $\angle p_1 p_0 \tilde{q}_1$ is an acute or right angle, as shown in Fig.~\ref{fig-Case-1}.

\hangafter=1 \hangindent=2.5em 
a) The points $p_0$, $p_1$($\tilde{q}_1$) and $p_2$($\tilde{q}_2$) are collinear as shown in Fig.~\ref{fig-Case-1}a. Then, $p_{0\thicksim 2}$ are termed as control points. 

\hangafter=1 \hangindent=2.5em 
b) Only $p_2$ is on $\overline{p_0 p_1}$ as shown in Fig.~\ref{fig-Case-1}b. 
Then the points $p_0$ and $q_{1\thicksim 3}^{\ast}$ are termed as control points, where $q_{1\thicksim 3}^{\ast}$ are determined by substituting the points $p_0$, $\tilde{q}_1$, $\tilde{q}_2$($p_1$) and $\tilde{q}_3$ into the function \textit{OptCubic} defined as follows, where $\tilde{q}_3$ is on the ${p_1 p_2}$ segment and is determined to make the convex hull maximal and collision-free.

\hangafter=1 \hangindent=2.5em 
c) Only $\tilde{q}_1$ is on $\overline{p_{0}p_{1}}$ as shown in Fig.~\ref{fig-Case-1}c.
Then the points $p_0$, $p_1$($\tilde{q}_1$), and $q_2^{\ast}$ are termed as control points, where $q_2^{\ast}$ is determined by substituting the points $p_{0\thicksim 1}$ and $\tilde{q}_2$ into the function \textit{OptQuad1} defined as follows, where $\tilde{q}_2$ is on the ${p_1 p_2}$ segment and is chosen to make the convex hull maximal and collision-free. 

\noindent
\newline
\textit{(\textit{OptCubic}) Optimization of a cubic curve}: 
Define the input points as $p_{0\thicksim 3}$. Then the optimal points $p_{1\thicksim 3}^{\ast}$ can be obtained through $\min{|\kappa_3|_{\max}}$, which is to minimize a Bezier curve's maximum curvature.
To calculate $|\kappa_3|_{\max}$, the cubic curve $D$ with control points $p_{0\thicksim 3}$ is approximated by the two quadratic curves $E$ and $F$ with control points $e_{0\thicksim 2}$ and $f_{0\thicksim 2}$, respectively. The relationship of these control points is~\cite{choi2010real}:
\begin{equation*} \small
	\begin{cases}
		e_0 &= p_0 \\
		e_1 &= \frac{9}{32}p_0 + \frac{21}{32}p_1 + \frac{3}{32}p_2 - \frac{1}{32}p_3 \\
		e_2 &= \frac{1}{8}p_0 + \frac{3}{8}p_1 + \frac{3}{8}p_2 + \frac{1}{8}p_3 \\
		f_0 &= \frac{1}{8}p_0 + \frac{3}{8}p_1 + \frac{3}{8}p_2 + \frac{1}{8}p_3 \\
		f_1 & = -\frac{1}{32}p_0 + \frac{3}{32}p_1 + \frac{21}{32}p_2 + \frac{9}{32}p_3 \\
		f_2 &= p_3. \\
	\end{cases}
\end{equation*}

\begin{figure}[htb]
	\centering
	\begin{minipage}[p]{0.2\textwidth}
		\centerline{\includegraphics[height=3.7cm]{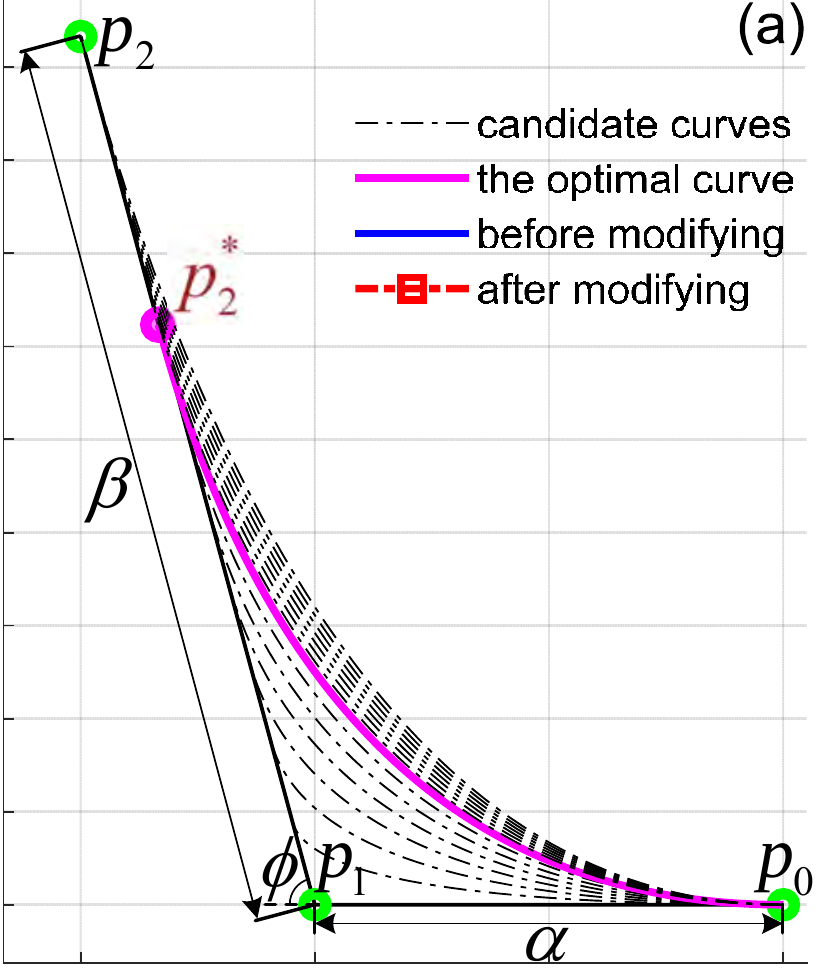}}
	\end{minipage}
	\quad
	\begin{minipage}[p]{0.2\textwidth}
		\centerline{\includegraphics[height=1.8cm]{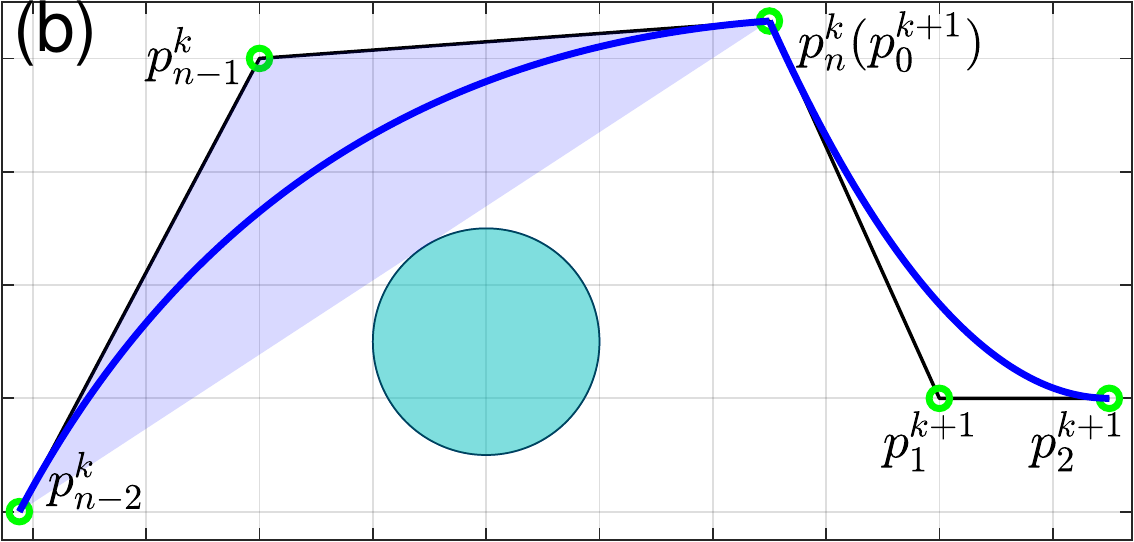}}
		\vfill
		\centerline{\includegraphics[height=1.9cm]{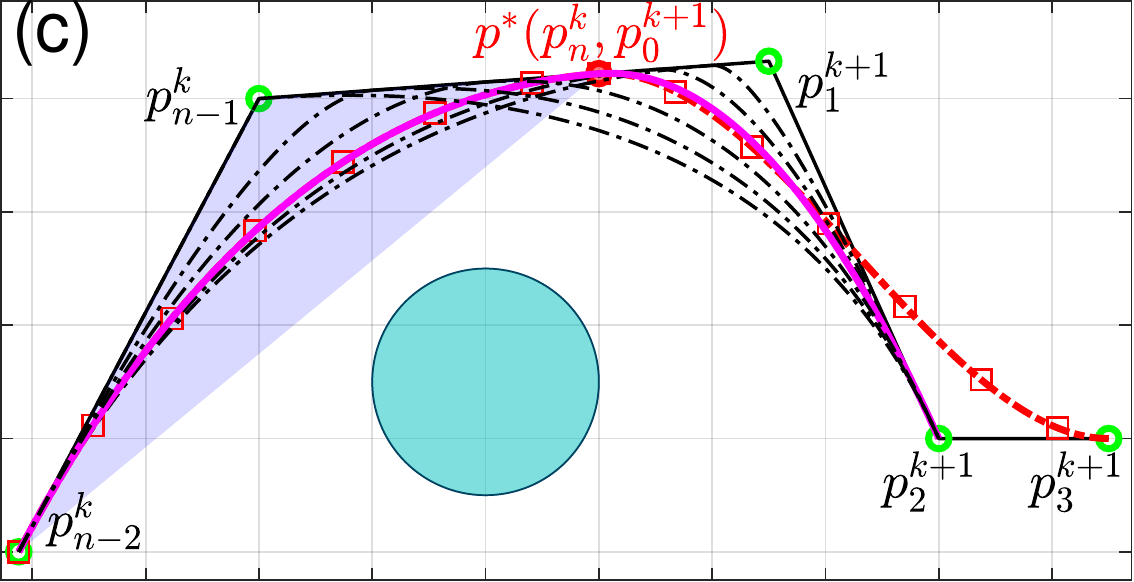}}
	\end{minipage}
	\caption{(a)The optimization of a quadratic curve with fixed values of $\alpha$ and $\phi$;(b, c)The optimization to make the curve $C^{1}$ continuous.}
	\label{fig-opt-one-midpoint} 
\end{figure}

\noindent
\newline
\textit{(\textit{OptQuad1}) Optimization of a quadratic curve}: 
Define the input control points as $p_{0\thicksim 2}$. Then the optimal point $p_{2}^{\ast}$ can be determined by $\min{|\kappa_2|_{\max}}$. That is to make length of the $p_1 p^{\ast}_2$ segment equal to $\min{(\beta,\frac{-\cos{\phi}+\sqrt{\cos^2(\phi)+8}}{2}\alpha)}$~\cite{choi2010real} when the values of $\phi$ and $\alpha$ are fixed, where $\phi$, $\alpha$, and $\beta$ denote the included angle of $\overrightarrow{p_0 p_1}$ and $\overrightarrow{p_1 p_2}$, and length of the $p_0 p_1$ and $p_1 p_2$ segments, respectively, as shown in Fig.~\ref{fig-opt-one-midpoint}a.

\begin{figure}[htb]
	\centering
	\begin{minipage}[p]{0.23\textwidth}
		\centerline{\includegraphics[height=2cm]{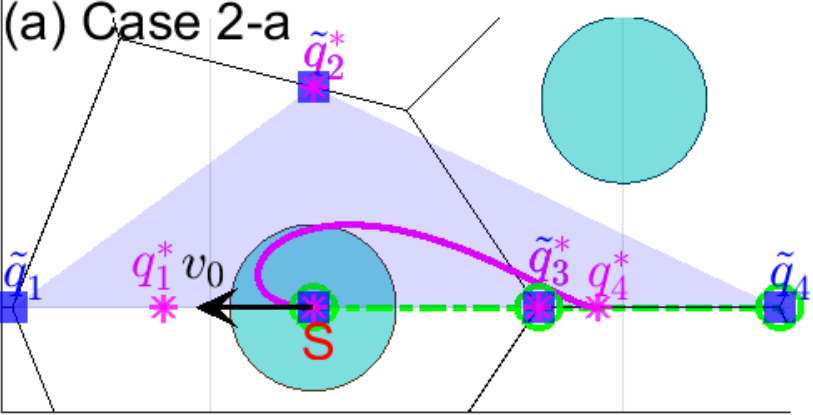}} 
		\vfill  
		\centerline{\includegraphics[height=2cm]{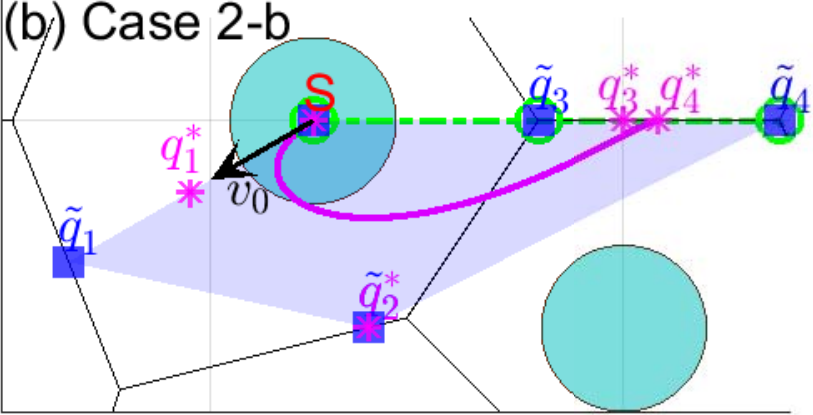}}   
	\end{minipage}
	\begin{minipage}[p]{0.23\textwidth}
		\centerline{\includegraphics[height=4cm]{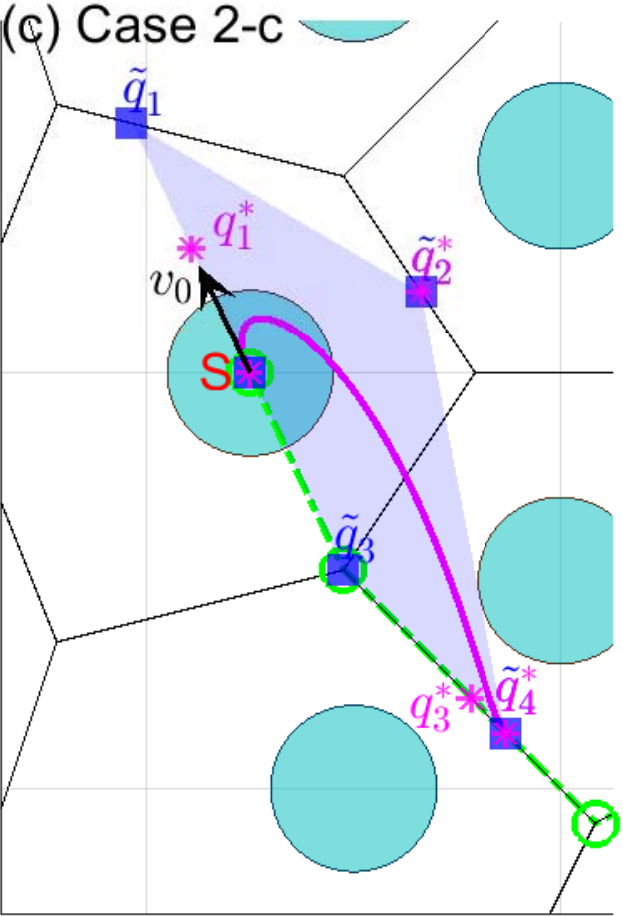}}   
	\end{minipage}
	\caption{Case 2: The three subgraphs correspond to Case 2-a, Case2-b and Case 2-c, respectively.} 
	\label{fig-Case-2} 
\end{figure}

\hangafter=1 \hangindent=3.5em \noindent
Case 2: The point $\tilde{q}_1$ and/or $p_2$ is on $\overline{p_0 p_1}$, and the angle $\angle p_1 p_0 \tilde{q}_1$ is an obtuse or straight angle, as in Fig.~\ref{fig-Case-2}.

\hangafter=1 \hangindent=2.5em 
a) The points $p_0$, $\tilde{q}_1$, $p_1$ and $p_2$ are collinear as in Fig.~\ref{fig-Case-2}a. Then the points $p_0$ and $q_{1\thicksim 4}^{\ast}$ are termed as control points, where $q_{1\thicksim 4}^{\ast}$ are determined by substituting the points $p_0$, $\tilde{q}_1$, $\tilde{q}_2$, $\tilde{q}_3$($p_1$) and $\tilde{q}_4$ into the function \textit{OptQuar} defined as follows, where $\tilde{q}_4$ is on the ${p_1 p_2}$ segment and is chosen to make the convex hull maximal and collision-free, and $\tilde{q}_2$ is the point farther away from $p_0$ among the two intersection points of the convex cell and the line perpendicular to $\overline{p_0 p_1}$ across $p_0$.

\hangafter=1 \hangindent=2.5em 
b) Only $p_2$ is on $\overline{p_{0}p_{1}}$ as shown in Fig.~\ref{fig-Case-2}b.
Then the selection strategy is similar to that in Case2-a. The only difference is that $\tilde{q}_2$ is selected such that $\overrightarrow{p_0 \tilde{q}_2}$ is the interior bisector of the angle $\angle p_1 p_0 \tilde{q}_1$.

\hangafter=1 \hangindent=2.5em 
c) Only $\tilde{q}_1$ is on $\overline{p_{0}p_{1}}$ as shown in Fig.~\ref{fig-Case-2}c.
Then the selection strategy is similar to that in Case2-a. The only difference is that $\tilde{q}_2$ is selected such that $\overline{p_0 \tilde{q}_2}$ is perpendicular to $\overline{p_0 p_1}$ and is inside the angle $\angle p_0 p_1 p_2$.

\noindent
\newline
\textit{(\textit{OptQuar}) Optimization of a quartic curve}:  
Define the input points as $p_{0\thicksim 4}$. Then the optimal points $p_{1\thicksim 4}^{\ast}$ can be determined by $\min{|\kappa_4|_{\max}}$.
To calculate $|\kappa_4|_{\max}$, the following strategies are designed. A quartic curve $D$ with control points $p_{0\thicksim 4}$ is first approximated by two cubic curves $E$ and $F$ with control points $e_{0\thicksim 3}$ and $f_{0\thicksim 3}$, respectively. Then combining the \textit{OptCubic} function, the value of $\min{|\kappa_4|_{\max}(D)}$ can be obtained. The relationship of these control points is:
\begin{equation*} \small
	\begin{cases}
		e_0 &= p_0 \\
		e_1 &= \frac{1}{672}(227p_0 + 436p_1 + 18p_2 - 12p_3 + 3p_4) \\
		e_2 &= \frac{1}{672}(101p_0 + 268p_1 + 270p_2 + 44p_3 - 11p_4) \\
		e_3 &= \frac{1}{16}(p_0 + 4p_1 + 6p_2 + 4p_3 + p_4) \\
		f_0 &= \frac{1}{16}(p_0 + 4p_1 + 6p_2 + 4p_3 + p_4) \\
		f_1 &= \frac{1}{672}(44p_1 - 11p_0 + 270p_2 + 268p_3 + 101p_4) \\
		f_2 &= \frac{1}{672}(3p_0 - 12p_1 + 18p_2 + 436p_3 + 227p_4) \\
		f_3 &= p_4.  \\
	\end{cases}
\end{equation*}

\begin{figure}[htb]
	\centering
	\begin{minipage}[t]{0.2\textwidth}
		\centerline{\includegraphics[height=4cm]{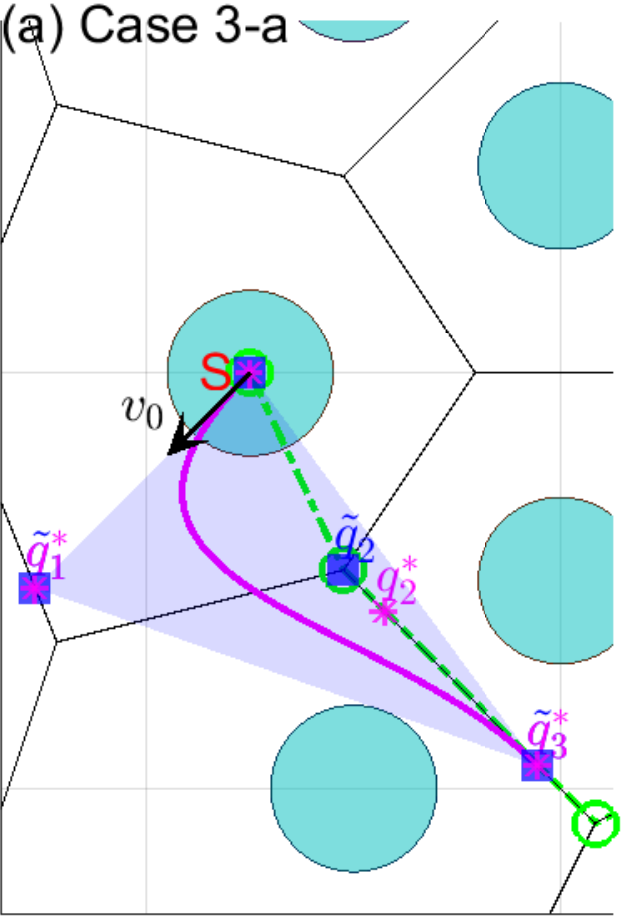}}   
	\end{minipage}
	\begin{minipage}[t]{0.2\textwidth}
		\centerline{\includegraphics[height=4cm]{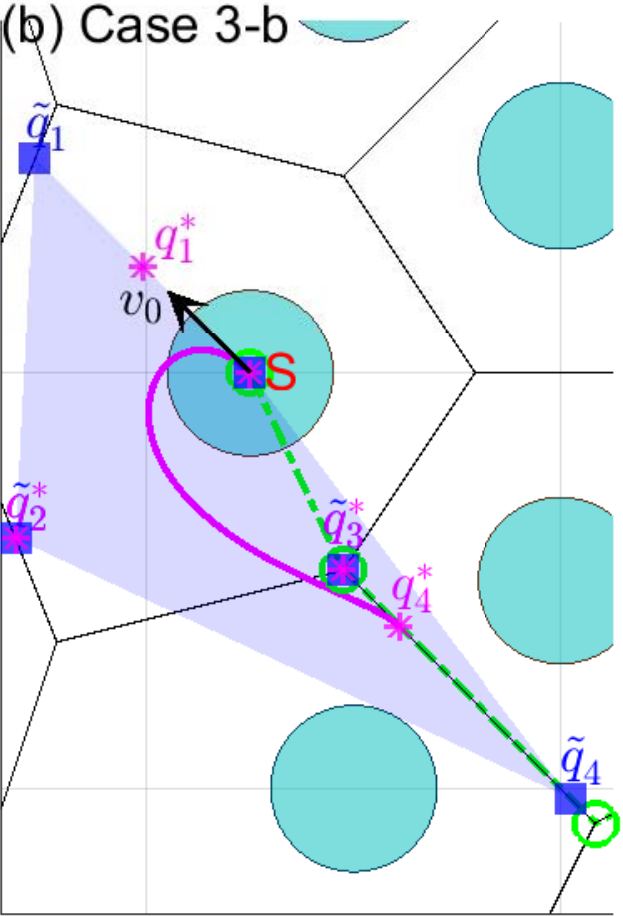}}   
	\end{minipage}
	\caption{Case 3: The two subgraphs correspond to Case3-a and Case3-b, respectively.} 
	\label{fig-Case-3} 
\end{figure}

\hangafter=1 \hangindent=3.5em \noindent
Case 3: Neither $\tilde{q}_1$ nor $p_2$ is on $\overline{p_{0}p_{1}}$, and the points $\tilde{q}_1$ and $p_2$ are on the different side of $\overline{p_{0}p_{1}}$, as shown in Fig.~\ref{fig-Case-3}.

\hangafter=1 \hangindent=2.5em 
a) The angle $\angle p_1 p_0 \tilde{q}_1$ is an acute or right angle as shown in Fig.~\ref{fig-Case-3}a.
Then the selection strategy is similar to that in Case1-b.

\hangafter=1 \hangindent=2.5em 
b) The angle $\angle p_1 p_0 \tilde{q}_1$ is an obtuse or straight angle as shown in Fig.~\ref{fig-Case-3}b.
Then the selection strategy is similar to that in Case2-b.

\begin{figure}[htb]
	\centering
	\begin{minipage}[t]{0.155\textwidth}
		\centerline{\includegraphics[height=4cm]{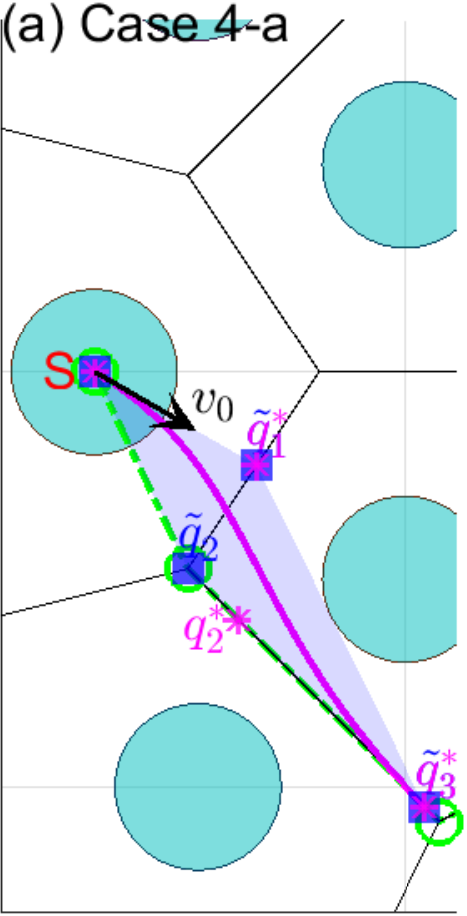}}   
	\end{minipage}
	\begin{minipage}[t]{0.155\textwidth}
		\centerline{\includegraphics[height=4cm]{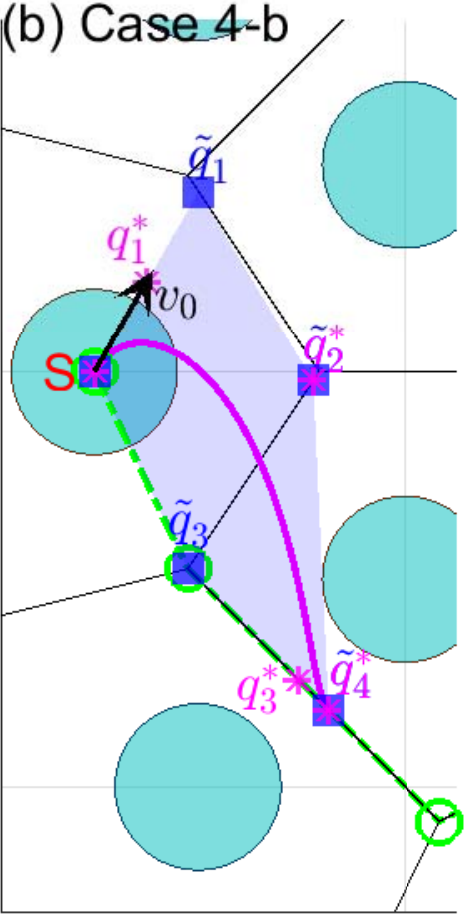}}   
	\end{minipage}
	\begin{minipage}[t]{0.155\textwidth}
		\centerline{\includegraphics[height=4cm]{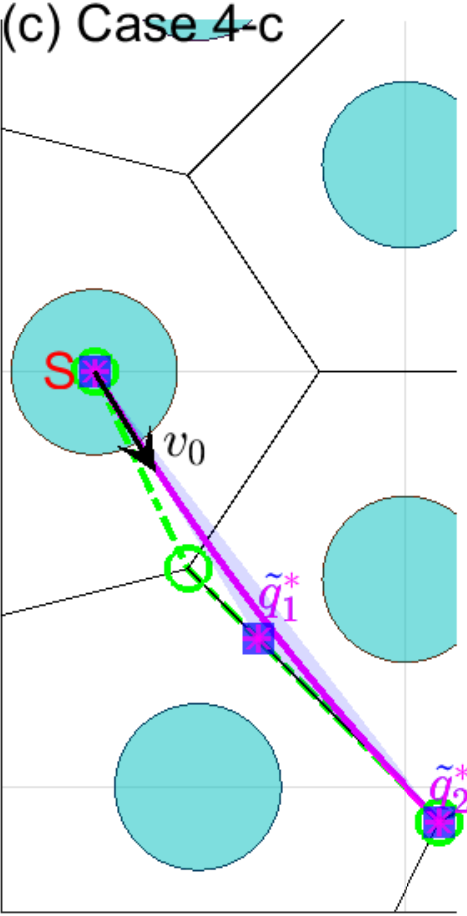}}   
	\end{minipage}
	\caption{Case 4: The three subgraphs correspond to Case 4-a, Case 4-b and Case 4-c, respectively.} 
	\label{fig-Case-4} 
\end{figure}

\hangafter=1 \hangindent=3.5em \noindent
Case 4: Neither $\tilde{q}_1$ nor $p_2$ is on $\overline{p_{0}p_{1}}$, and the points $\tilde{q}_1$ and $p_2$ are on the same side of $\overline{p_{0}p_{1}}$, as shown in Fig.~\ref{fig-Case-4}.

\hangafter=1 \hangindent=2.5em 
a) There is no intersection of the segments ${p_0 \tilde{q}_1}$ and ${p_1 p_2}$ as shown in Fig.~\ref{fig-Case-4}a, or they are intersecting but the ${p_0 q_1}$ segment collides with obstacles as shown in Fig.~\ref{fig-Case-4}b. 
Then the selection strategy is similar to that in Case1-b, if the angle $\angle p_1 p_0 \tilde{q}_1$ is an acute or right angle, otherwise the strategy will be similar to that in Case2-b.

\hangafter=1 \hangindent=2.5em 
b) The two segments ${p_0 \tilde{q}_1}$ and ${p_1 p_2}$ are intersecting, and the ${p_0 \tilde{q}_1}$ segment is collision-free as shown in Fig.~\ref{fig-Case-4}c.
Then the selection strategy is similar to that in Case1-c. The only difference is that $p_1$ and $\tilde{q}_1$ are not the same point, and $\tilde{q}_1$ is termed as a control point.

\begin{figure}[htb]
	\centerline{\includegraphics[width=2.5in,height=1.545in]{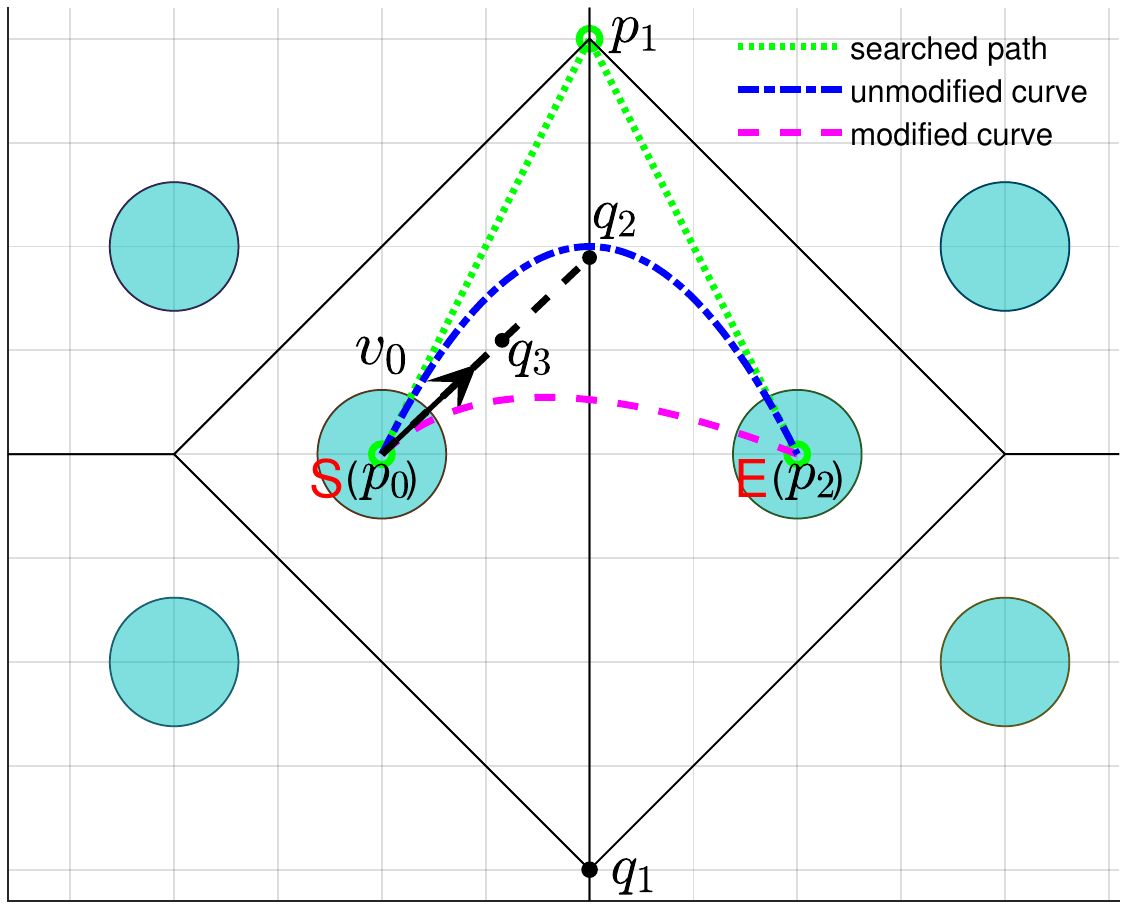}}   
	\caption{The case of environment with sparse obstacles.} 
	\label{fig-purpose-sparse} 
\end{figure}

\textbf{2) Design the first B{\'e}zier curve for the environment with sparse obstacles}: 

There will be some problems, if environmental obstacles are sparse as shown in Fig.~\ref{fig-purpose-sparse}, and Voronoi vertices are directly set as control points. From this figure, we see that there are big jerks and unnecessary routes in the obtained green path. To address this kind of situation, the second waypoint $p_1$ is replaced by a new point. The method for determining the new point is described as follows.  

If there just exist three waypoints (i.e., $p_{0\thicksim 3}$) in the obtained green path as shown in Fig.~\ref{fig-purpose-sparse}, and the cells of starting and end points have the public edge $\overline{p_1 q_1}$, and the middle point of the $p_0 p_2$ segment is on $\overline{p_1 q_1}$, and there is an intersection point $q_2$ of the edge $\overline{p_1 q_1}$ and the extension of velocity vector $\overrightarrow{p_0 q_2}$, then the second waypoint $p_1$ can be replaced by the new point $q_3$ which is on $\overline{p_0 q_2}$. The length of the $p_0 q_3$ segment is determined by substituting the points $p_0$, $q_3$, $q_2$ and $p_2$ into the function \textit{OptQuad1}. Fig.~\ref{fig-purpose-sparse} shows the modified purple curve is less conservative than the unmodified blue one.

\textbf{Step 2: Smooth the remaining B{\'e}zier curves}

\begin{figure}[htb]
	\centerline{\includegraphics[width=2.5in]{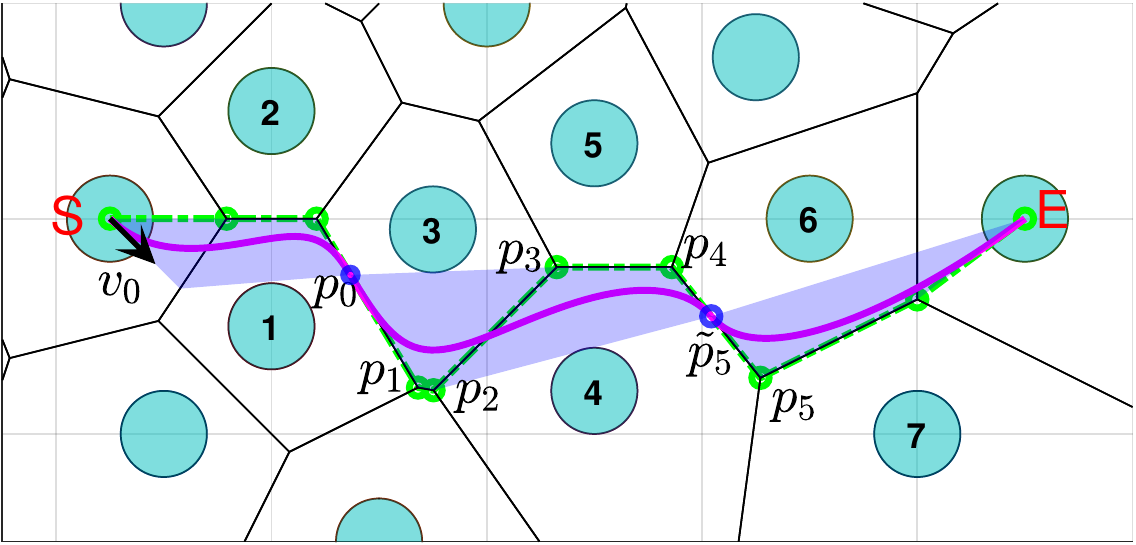}}   
	\caption{The selection of remaining control points.} 
	\label{fig-collision-free} 
\end{figure}

The method for determining remaining curves is demonstrated in Fig.~\ref{fig-collision-free}.
When we add $p_5$ into the current point set including $p_{0\sim 4}$, the convex cell constructed by $p_{0\sim 5}$ collides with the obstacle labeled `4'. 
To solve this, $\tilde{p}_5$ on the $p_4 p_5$ segment is determined to make the convex cell constructed by $p_{0\sim 4}$ and $\tilde{p}_5$ collision-free and maximal. 


The obtained piecewise curve will not be $G^1$ continuous at very few special connection points which are the last control points of corresponding curves. To be $G^1$ continuous, such control points are modified by substituting the last three control points of current curve and the first two control points of next curve into the function \textit{OptQuad2} defined as:

\noindent
\textit{(\textit{OptQuad2}) Optimization of a quadratic curve}: 
Define the input points $p_{(n-2)\thicksim n}^{k}$ and $p_{0 \thicksim 1}^{k+1}$ as control points of the curves $D$ and $E$, respectively. As shown in Fig.~\ref{fig-opt-one-midpoint}b, the waypoint $p_n^k$ happens to be the join point, where the piecewise curve has $G^0$ continuity. To be $G^1$ continuous, the $p_{n}^{k}$ of curve $D$ is replaced by the new point $\tilde{p}$, and $\tilde{p}$ is also added to be the first control point of curve $E$, where $\tilde{p}$ is chosen on the $p_{n-1}^k p_n^{k}$ segment. The optimal position of $\tilde{p}$ is determined by $\min{\{|\kappa_2|_{\max}(D),|\kappa_2|_{\max}(E)\}}$. Combining the \textit{OptQuad1} function, the optimal point $p^{\ast}$ is obtained as shown in Fig.~\ref{fig-opt-one-midpoint}c.

\subsection{Transform paths to trajectories}
A B{\'e}zier curve can be mathematically described as:
\begin{equation} \small \label{eq_bezier}
	P(\theta)=\sum_{i=0}^{n}p_{i}B_{i,n}(\theta), \theta \in [0,1], 
\end{equation}
where $P(\cdot)\triangleq [P_x(\cdot), P_y(\cdot)]^T$ denotes the $n$-order B{\'e}zier curve with control points $p_{0\thicksim n}$, and $B_{i,n}(\theta) \triangleq C_{n}^{i}\theta^{i}(1-\theta)^{n-i}$. Then, the piecewise B{\'e}zier curve can be derived by substituting the sets of control points obtained in Section~\ref{sec_smooth} into~\eqref{eq_bezier}. As a result, the time-dependent reference can be formulated as: 
$\bm{x}_d(t)$=$[ P_x(\theta(t))$, $P_y(\theta(t))$, $\atantwo(\dot{P}_y(\theta)\dot{\theta}(t), \dot{P}_x(\theta)\dot{\theta}(t))$, $u(\theta)\dot{\theta}(t)$, 0, $r(\theta) \dot{\theta}(t) ]^T$,
where $u(\theta)=\|\dot{P}(\theta)\|$, and $r(\theta)=(\dot{P}_x(\theta) \ddot{P}_y(\theta)-\dot{P}_y(\theta) \ddot{P}_x(\theta))/\|\dot{P}(\theta)\|^2$.

From the expression of $\bm{x}_d(t)$, it is clear that the speed profile $\dot{\theta}(t)$ should be designed. Here, a simple but helpful scheme is designed~\cite{fossen2011handbook}:
\begin{equation} \small \begin{cases}
		\dot{\theta}(t)={u_{d}(t)} / {\|\dot{P}(\theta)\|} \\
		T_{\theta}\dot{u}_{d}(t)+u_{d}(t)=U_{d},\\
\end{cases} \end{equation}
where $u_{d}(t)$, $U_{d}$ and $T_{\theta}$ stand for the reference surge velocity, the desired velocity, and a time constant respecting physical limitations of marine vehicles, respectively. 

\section{EMPC-based planner design} \label{sec_alg_des}
In this paper, the planned motion ($\bar{\bm{x}}(s)$, $\bar{\bm{\tau}}(s)$), $s \in{[t_k,t_k+T_p]}$,  at time instant $t_k$ is obtained by solving the following EMPC problem, $\min{J(\cdot)}$.
\begin{problem} \small  \label{pro_1}
	\begin{align}
		&\min_{\bar{\bm{\tau}}_{\delta}(s)}\int_{t_k}^{t_k+T_p}(l_{e}(\bar{\bm{\tau}}, s)+l_{t}(\bar{\bm{x}},\bm{x}_d, s)+ \|\bar{\bm{\tau}}_{\delta}(s)\|_{R_{\delta}}^2 \nonumber \\
		&\hspace{2.25cm}+ \|\bar{\bm{\tau}}(s)\|_{R_u}^2 )ds + m(\bar{\bm{x}}, \bm{x}_d, t_k+T_p)  \label{eq_integral} \nonumber \\
		&s.t. ~\dot{\bar{\bm{x}}}_{\text{aug}}(s)=f_{\text{aug}}(\bar{\bm{x}}_{\text{aug}}(s),\bar{\bm{\tau}}_{\delta}(s)), s \in [t_k,t_k+T_p], \nonumber \\
		&~~~~~~\bar{\bm{x}}_{\text{aug}}(s) \in \mathcal{X}_{\text{aug}}, \bar{\bm{\tau}}_{\delta}(s) \in \mathcal{U}_{\delta}, s \in [t_k,t_k+T_p], \nonumber \\
		&~~~~~~c_{obs}(\bar{\bm{x}}_p(s),\bm{x}_o^i) \geq 0, s \in [t_k,t_k+T_p], \nonumber
	\end{align}
\end{problem}
where $l_e(.)$ is the economic term specifying the power consumption; $l_t(.,.)$ is the tracking term representing the trajectory tracking error, in which $\bm{x}_d$ is the reference trajectory; $m(.,.)$ is the terminal penalty function; $R_{\delta}$ and $R_u$ are weighting matrices to penalize the motion's jerk and acceleration, respectively;  $\mathcal{X}_{\text{aug}}$ and $\mathcal{U}_{\delta}$ are constraint sets on state and input, respectively; $c_{obs}$ describes collision-free conditions; $\bm{x}_o^i$ is the center point of the detected obstacle $i$; $T_p$ denotes the prediction horizon. 






In what follows, we specify the detailed cost terms of the EMPC in Problem \ref{pro_1}.

\textbf{The economic term}: Using the exponential relationship between thrust force and electrical power~\cite{estrada2018forceful}, a cost function estimating the energy consumed by vessel's two propellers is constructed as: 
\begin{equation} \small \label{eq_le}
	\begin{aligned}
		l_e(\bm{\tau}, t) &\triangleq {k_e} ( {k_c}|F_l(t)|^{1.5}+{k_c}|F_r(t)|^{1.5})\\
		&= k_{ec} ( |F_l(t)|^{1.5} + |F_r(t)|^{1.5}),
	\end{aligned}
\end{equation}
where $F_l(t)\triangleq 0.5(X(t)+N(t)/d)$ and $F_r(t)\triangleq 0.5(X(t)-N(t)/d)$ are the thrust force of left and right propellers, respectively; $k_{ec}\triangleq {k_e} {k_c}$; $k_e$, $k_c$ and $d$ are the weighting factor, the constant coefficient, and the moment arm in yaw, respectively. \eqref{eq_le} represents the physical power of actuators, which means the total consumed energy measured by Joules can be obtained by the integration of~\eqref{eq_le}.

\textbf{Safety tracking term}: In most cases, we also need marine vehicles to move close to the safety trajectories. As such, the penalty of deviation from the reference is designed as:
\begin{equation} \small \label{eq_lt}
	l_t(\bm{x}, \bm{x}_d, t) \triangleq \|\bm{x}_e(t)\|_Q^2 = \|\bm{x}(t) - \bm{x}_d(t)\|_{Q}^2,
\end{equation}
where $Q$ is a weighting matrix, $\bm{x}_e \triangleq [x_e,y_e,\psi_e,u_e,v_e,r_e]^T$, and $\bm{x}_d(t)\triangleq [x_d,y_d,\psi_d,u_d,v_d,r_d]^T$.

\textbf{Terminal cost}: The imposed terminal cost is designed as:  
\begin{equation} \small \label{eq_soft_ter}
	m(\bm{x}, \bm{x}_d, t) \triangleq \|\bm{x}(t) - \bm{x}_d(t)\|_P^2,
\end{equation} 
where $P$ is a weighting matrix, and $\bm{x}_d(t)$ is the reference trajectory for tracking.

\textbf{Collision-free conditions}:
To satisfy collision-free conditions, the following hard constraints are imposed:
\begin{equation} \small \label{eq_col_0}
	c_{obs}(\bm{x}_p(t),\bm{x}^i_o) \triangleq \|\bm{x}_p(t) - \bm{x}_o^i\|^2 - (r_c+r_v+r_o)^2 \geq 0,
\end{equation} 
where $r_c$, $r_{v}$, and $r_o$ are the clearance distance, and the radii of vehicle and obstacle, respectively.

%

%
Note that Problem~\ref{pro_1} has two new features in comparison with the conventional MPC-based motion planning method: An economic term is proposed in the stage cost, which can explicitly consider the physical power consumption in planning; the trajectory $\bm{x}_d$ with safety margin is designed as the state tracking reference.

\section{Algorithm design and implementation} \label{sec_online}
In this section, the detailed receding horizon motion planning and reference construction trajectory by considering the sensor range and tracking error in practice are presented. 


\subsection{Receding horizon planning strategy}\label{sec_receding}
In most cases, the calculation time for planning is not explicitly incorporated into the planning process. As a result, the performance of planned motion will degrade as the calculation time increases.
To deal with this problem, a receding horizon planning scheme considering the time consuming for planning is proposed. 

\begin{figure}[htb]
	\centerline{\includegraphics[width=2.5in]{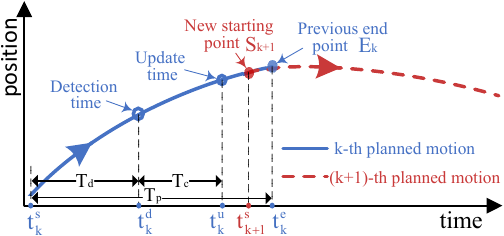}}   
	\caption{The proposed strategy of receding horizon planning.}  
	\label{fig-horizon-plan}  
\end{figure}

The detailed planning process is shown in Fig.~\ref{fig-horizon-plan}. Assume a vehicle is starting to execute the $k^{th}$ planned motion with length $T_p$ at time $t_k^s$. After $T_d$ seconds, the vehicle starts to detect the environment and to plan a new motion starting from $t_{k+1}^s$ which is between the time period of $t_k^u$ and $t_k^e$. After $T_c$ seconds(which is the computation time), the planner generates the new planned motion for the next time step $t_{k+1}^s$. At time $t_{k+1}^s$, the above step repeat.




\subsection{Construction of the reference trajectory} \label{sec_const_ref}
To accommodate the receding horizon planning strategy, the reference trajectory should be periodically replanned. At this point, two issues arise, namely, where the new reference starts, and how to sequentially construct a safe reference. 

\begin{figure}[htb]
	\centerline{\includegraphics[width=2.5in, height = 1.44in]{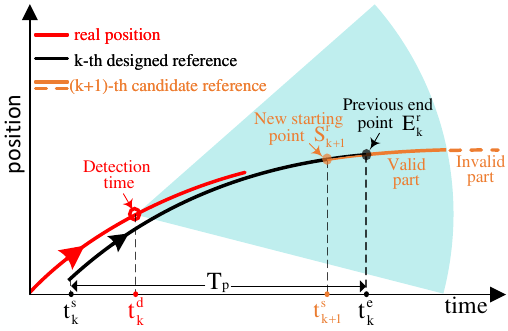}}   
	\caption{The proposed strategy of constructing references.}  
	\label{fig-horizon-plan2}  
\end{figure}

For the first issue, if the reference is replanned from vehicle's current position, the energy-saving effect will be very limited due to the trade off between the tracking and economic terms in Problem~\ref{pro_1}. To achieve a greater degree of energy conservation, and better match the strategy, the new starting point `$\text{S}_{k+1}^r$' of reference trajectory is chosen to be the position at time $t_{k+1}^s$ on $k^{th}$ generated reference (see Fig. \ref{fig-horizon-plan2} ). In this way, there is much room for tracking term to balance the economic term, which means the better energy-saving effect can be obtained.

To deal with the second issue, the vehicle will periodically generate a new reference from `$\text{S}_{k+1}^r$' with length $T_p$ at time $t_k^d$. If some of the new reference is not within the region detected at $t_k^d$ (shadow area shown in Fig.~\ref{fig-horizon-plan2}), the reference is not safe and can only be regarded as a candidate reference. 
Then a safe one is constructed by combining the last unexectued part of the previous reference and the candidate reference, which means only the valid part of the candidate reference (i.e., the part within the detected region) is selected.

\subsection{Online motion planning algorithm}
Based on Problem~\ref{pro_1} and the design of its elements, the receding horizon motion planning algorithm is formulated in Algorithm~\ref{alg_2}.



\begin{algorithm}[htb] \small
	\begin{algorithmic}[1]
		\caption{Generate the motion}
		\label{alg_2}	
		\Require Weighting factors $Q$, $P$, $R_{\delta}$ and $R_u$, parameters $k_e$ and $k_c$, prediction horizon $T_p$, constraint sets $\mathcal{X}_{\text{aug}}$ and $\mathcal{U}_{\delta}$, and time constants $T_c$ and $T_d$;
		\While{task not completed}
		\If{$t = t_k^d$ ($t = 0$ for the first time)} 
		\State $\bm{x}_{\text{aug}}(t_{k+1}) \leftarrow \bm{x}_{\text{aug}}^{\ast}(t_{k+1}^s)$; 
		\State Update $c_{obs}(\bar{\bm{x}}_p(s),\bm{x}_o^i) \geq 0, s \in [t_{k+1},t_{k+2}]$;
		\State $\bm{x}_{d[t_{k+1},t_{k+2}]}\leftarrow $GenerateRef($\bm{x}_{d[t_{k},t_{k+1}]},\hat{\bm{x}}_{d[t_{k+1},t_{k+2}]})$;
		\State $\bm{x}_{\text{aug}[t_{k+1},t_{k+2}]}^{\ast}\leftarrow$Solve $\min{J(\bm{x}_{\text{aug}}(t_{k+1}),t_{k+1},t_{k+2})}$;
		\ElsIf{$t = t_k^u$ ($t = T_c$ for the first time)}
		\State $\bm{x}^{\ast}_{\text{aug}[t_k^u,t_k^u+2T_p-T_d-T_c]} \leftarrow$[$\bm{x}^{\ast}_{\text{aug}[t_k^u,t_k^s]}, \bm{x}_{\text{aug}[t_{k+1},t_{k+2}]}^{\ast}$)];
		\ElsIf{$t = t_k^s$}
		\State $k \leftarrow k+1$; 
		\EndIf		
		\EndWhile
	\end{algorithmic}
\end{algorithm}

\section{Experimental tests} \label{sec_hard_exp}
\begin{figure}[htb]
	\centerline{\includegraphics[width=2.5in, height=1.225in]{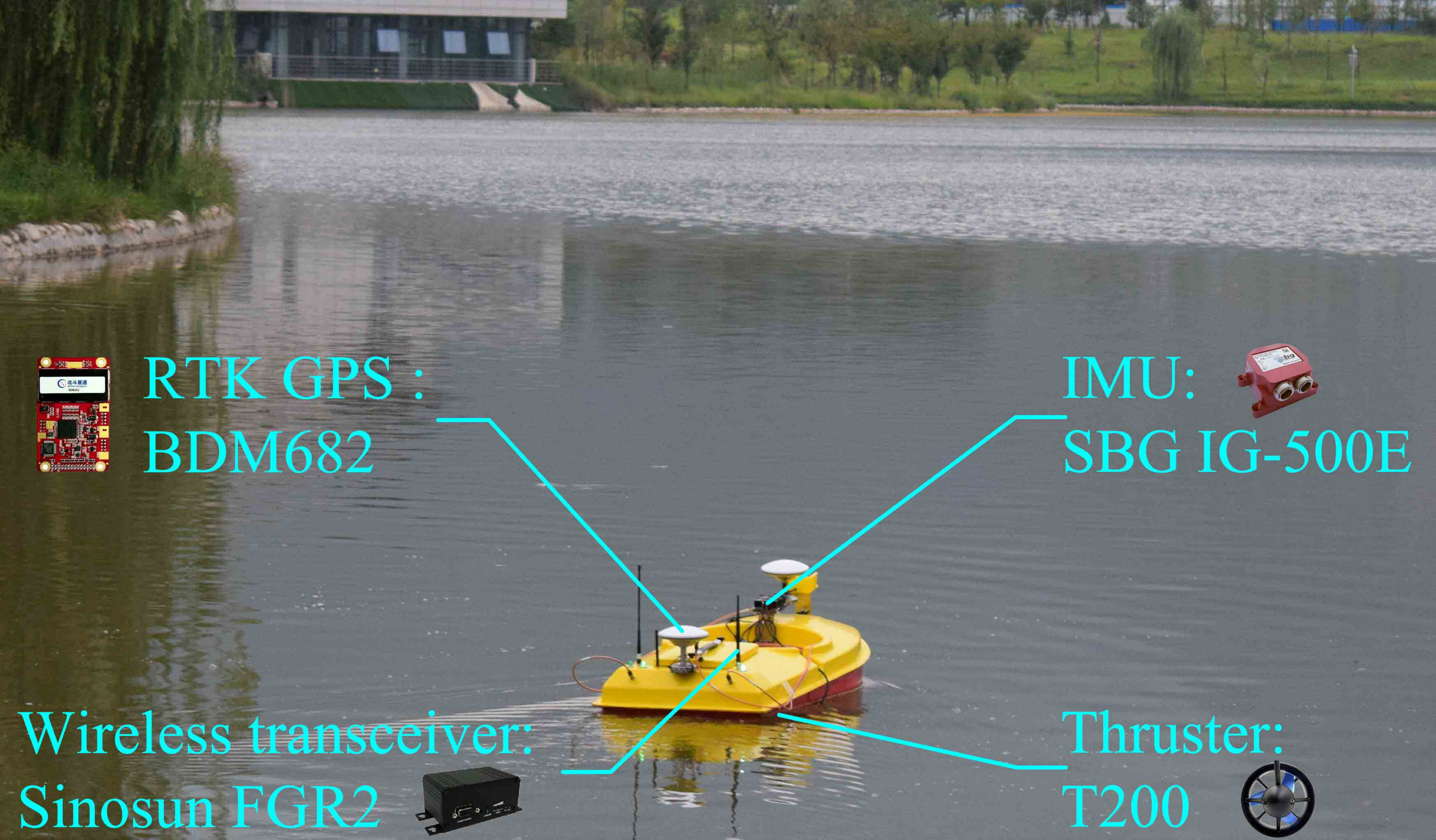}}    
	\caption{The tested mono-hull ship.} 
	\label{fig-vessel}                      
\end{figure}

To demonstrate the algorithm's feasibility and effectiveness in real-world scenarios, hardware experiments are conducted. The test platform is a  mono-hull ship shown in Fig.~\ref{fig-vessel}. 

For this vessel, its identified parameters are as follows: $M_1$=493.77, $M_2$=455.81, $M_3$=55.81, $D_1$=29.23, $D_2$=2173.7, $D_3$=17.7. The limitations of signals $X_{\delta}$, $N_{\delta}$, $X$ and $N$ are set as [-4.9, 4.9]~N/s, [-1.35, 1.35]~N$\cdot$m/s,  [-39.2, 39.2]~N and [-10.84, 10.84]~N$\cdot$m, respectively. The moment arm $d$, the distance $r_c$, and the radii $r_v$ and $r_o$ are set as 0.28 m, 0~m, 0.77~m, and 0.15~m, respectively. 

\subsection{Experimental settings} \label{sec_hard_setting}
In the planning task, the starting and end points are $[0,0]^T$ and $[5,60]^T$, respectively. To effectively compare the consumed energy for different strategies, the vehicle's initial state is set as $[0,0,90^{\circ},0,0,0]^T$. 
For the local planner, the sampling interval was set as 0.2~s, and the weighting matrices $Q$ = $P$ = diag(10,10,20,1,1,1), $R_{\delta}$ = diag(500,500), and $R_u$ = diag(0.1,0.1). The parameters are set as: $T_p$=20~s, $T_d$=13~s, $T_c$=2~s, and $U_{ref}$=0.2~m/s. The sensor's maximum detection range is set as 15~m, and the obstacles are placed virtually.

To track the planned motion in real-time, an MPC controller is utilized to track the planned motions, and its parameters are as follows: the prediction horizon and sampling period are set as 5~s and 0.2~s, respectively; the weighting matrices of stage state, control input, and terminal state are set as diag(10,10,0.5,0.1,0.1,0.1), diag($10^{-3}$,$10^{-3}$), and diag(100,100,5,0.1,0.1,0.1), respectively.

The parameter $k_c$ is identified as 0.95 according to the published raw data of T200 propellers~\cite{BlueRobotics2020}, which are equipped in the tested vessel. Besides, the optimization problems for planning and tracking are solved by the Gurobi solver~\cite{gurobi2019}. 

\subsection{Experimental results} \label{sec_hard_results}
\begin{figure}[htb]
	\centerline{\includegraphics[width=2.5in]{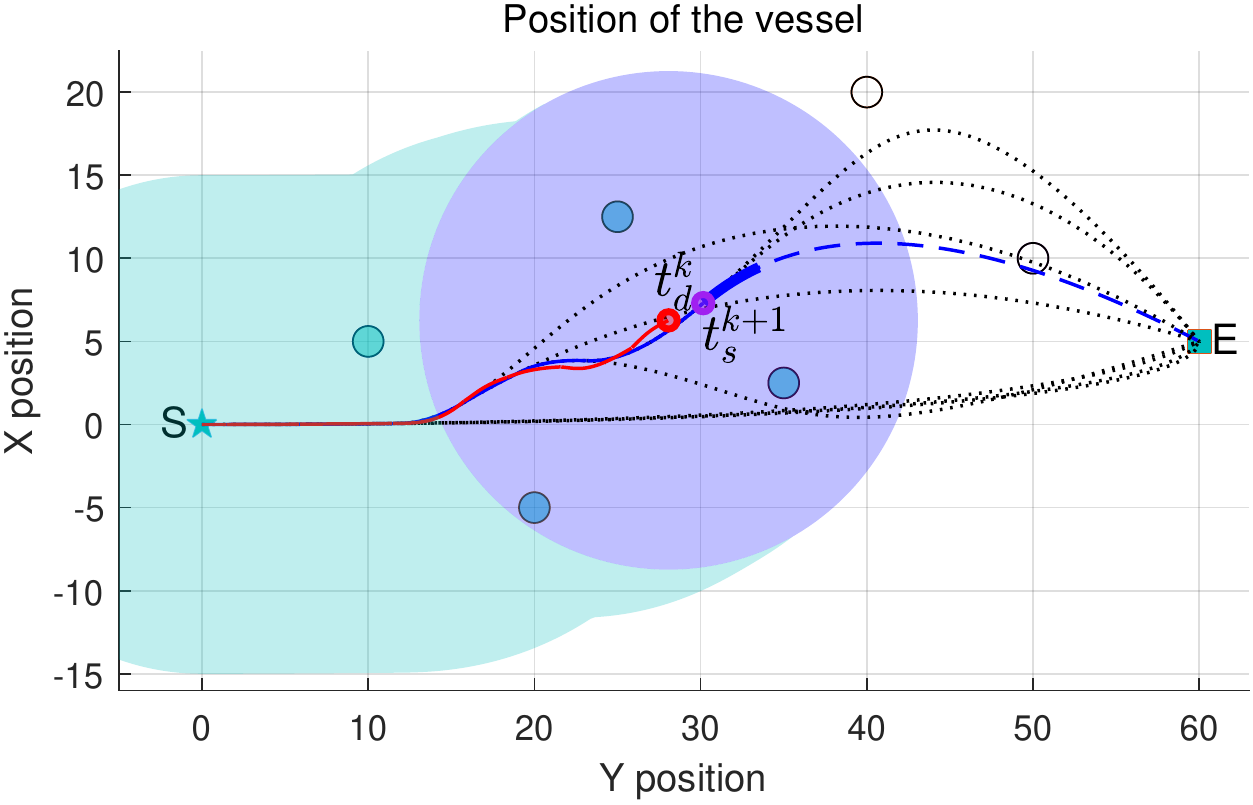}}    
	\caption{An illustration of the receding horizon planning process: The detected region consists of the light green area and the current covered region marked with the blue circle; small green and white circles denote the detected and undetected obstacles; the dotted lines show the trace of receding horizon planning; the red and the thin blue solid lines are the planned motion and reference; the dashed and the thick solid blue lines are the whole new planned trajectory and the part as reference.} 
	\label{fig-step}                      
\end{figure}

We first run the experiments for testing the feasibility of the method, and the results are shown in Fig.~\ref{fig-step}. From Fig.~\ref{fig-step}, it can be seen that the planned motion is collision-free and within the detection range.

\begin{figure}[htb]
	\centering{\includegraphics[width=2.5in]{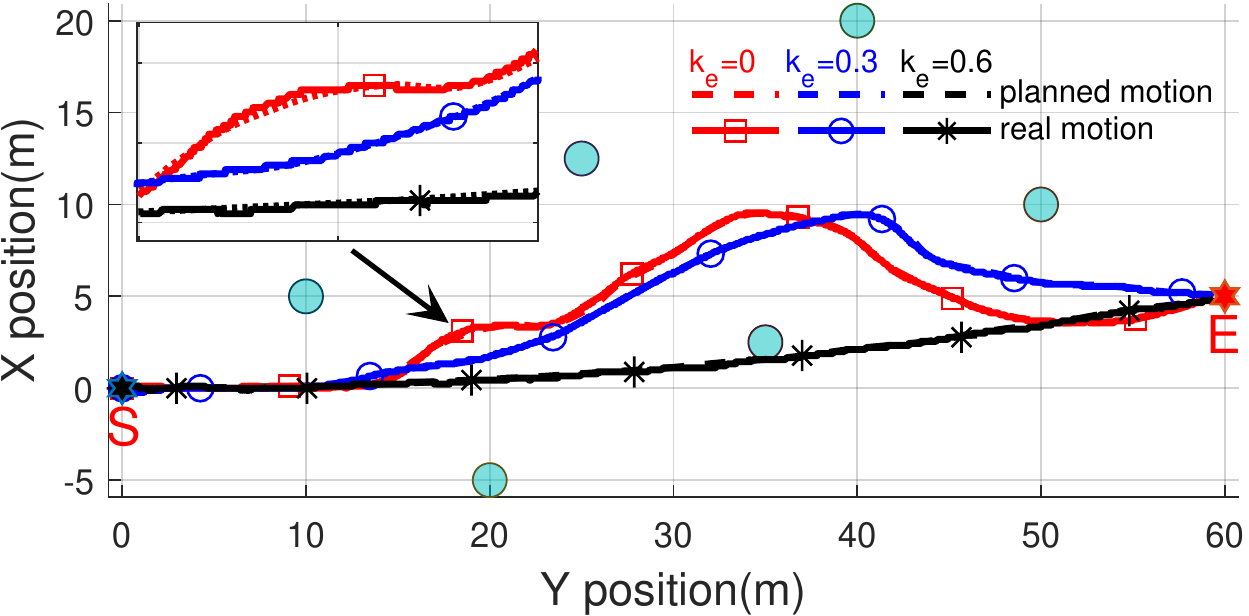}}    
	\caption{The overall planned motion and tracking results.} 
	\label{fig-position}                      
\end{figure}

\begin{figure}[htb]
	\centerline{\includegraphics[width=2.5in]{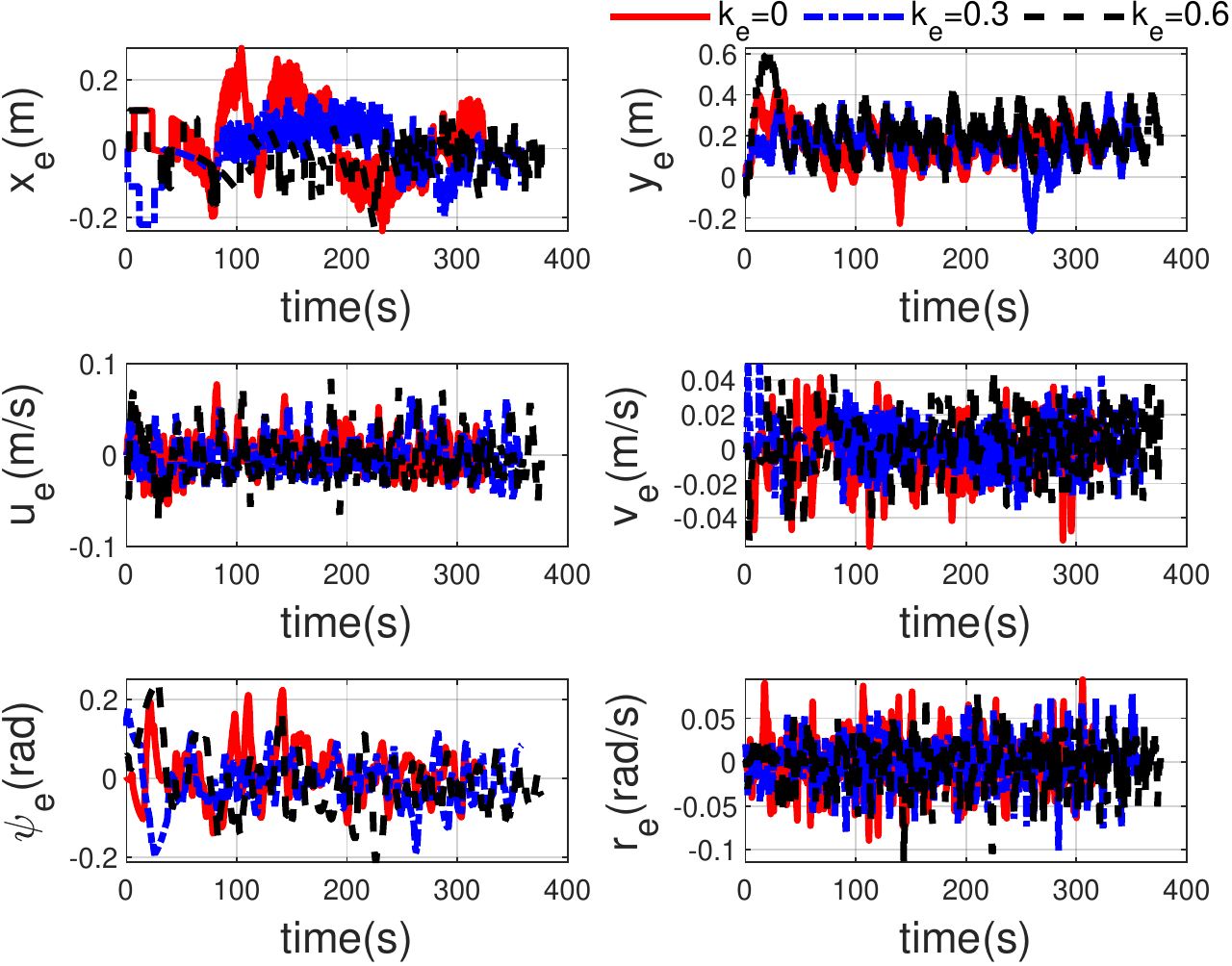}}    
	\caption{The tracking error.} 
	\label{fig-norm-part}                      
\end{figure}

To further show the tuning capability of the proposed method between energy consumption and safety margin, we run the experiments for three difference tuning parameter $k_{ec}$ in (\ref{eq_le}). The results are shown in Figs. \ref{fig-position}-\ref{fig-norm-part}. From these figures, it can be seen that the motion can be generated online, all the obstacles along the traveled route are detected (shown in Fig.~\ref{fig-position}), and the planned motion can be well tracked by the low-level controller (shown in Fig.~\ref{fig-norm-part}). 

In Table~\ref{tab-hard-consm}, we present the comparative results of power consumption with different values of the weighting factor $k_e$, where the effect of the economic term is demonstrated. It can be seen that with the increase of $k_e$, the amount of energy consumption declines and safety margin (distance to the obstacles) gradually increases, which well illustrates trade-off between the effects of the energy consumption and the safety margin.

\begin{table}[htb]
	\renewcommand{\arraystretch}{0.85}
	\caption{Power consumption with different settings.}
	\centering
	\label{tab-hard-consm}
	\setlength{\tabcolsep}{6mm}{ 
		\begin{tabular}{c c c c}
			\toprule
			\multicolumn{1}{c}{strategy} & \multicolumn{1}{c}{$k_{ec}$} & \multicolumn{1}{c}{consumption(J)}  \\ 
			\midrule
			\Rnum{1}  		&0     	&1103.86 \\
			\Rnum{2}		&0.3   	&1051.28 \\
			\Rnum{3} 		&0.6	&1025.33 \\			
			\bottomrule
	\end{tabular} }
\end{table}



\section{Conclusion} \label{sec_con} 
In this paper, we have designed a planning algorithm for marine vehicles in the form of receding horizon, where both vehicle dynamics and limitations are taken into account.
In particular, the proposed motion planning method is built an EMPC framework and can explicitly balance between safety margin and energy efficiency. 
Moreover, strategies on addressing the planning time and constructing safe references have been designed to facilitate the feasibility, safety, and implementation of planned motions in practice.
Finally, the effectiveness of the designed planning algorithm has been validated via hardware experiments.

\bibliographystyle{IEEEtran}
\bibliography{IEEEfull,planning}\ 

\begin{thebibliography}{10}
\providecommand{\url}[1]{#1}
\csname url@samestyle\endcsname
\providecommand{\newblock}{\relax}
\providecommand{\bibinfo}[2]{#2}
\providecommand{\BIBentrySTDinterwordspacing}{\spaceskip=0pt\relax}
\providecommand{\BIBentryALTinterwordstretchfactor}{4}
\providecommand{\BIBentryALTinterwordspacing}{\spaceskip=\fontdimen2\font plus
\BIBentryALTinterwordstretchfactor\fontdimen3\font minus
  \fontdimen4\font\relax}
\providecommand{\BIBforeignlanguage}[2]{{%
\expandafter\ifx\csname l@#1\endcsname\relax
\typeout{** WARNING: IEEEtran.bst: No hyphenation pattern has been}%
\typeout{** loaded for the language `#1'. Using the pattern for}%
\typeout{** the default language instead.}%
\else
\language=\csname l@#1\endcsname
\fi
#2}}
\providecommand{\BIBdecl}{\relax}
\BIBdecl

\bibitem{niu2018energy}
H.~Niu, Y.~Lu, A.~Savvaris\emph{,~et~al.}, ``An energy-efficient path planning
  algorithm for unmanned surface vehicles,'' \emph{Ocean Engineering}, vol.
  161, pp. 308--321, 2018.

\bibitem{chi2021generalized}
W.~Chi, Z.~Ding, J.~Wang\emph{,~et~al.}, ``A generalized {V}oronoi diagram
  based efficient heuristic path planning method for {RRT}s in mobile robots,''
  \emph{{IEEE} Transactions on Industrial Electronics}, 2021.

\bibitem{wu2021long}
G.~Wu, I.~Atilla, T.~Tahsin\emph{,~et~al.}, ``Long-voyage route planning method
  based on multi-scale visibility graph for autonomous ships,'' \emph{Ocean
  Engineering}, vol. 219, p. 108242, 2021.

\bibitem{khatib1986real}
O.~Khatib, ``Real-time obstacle avoidance for manipulators and mobile robots,''
  \emph{The International Journal of Robotics Research}, vol.~5, no.~1, pp.
  90--98, 1986.

\bibitem{huang2019motion}
Y.~Huang, H.~Ding, Y.~Zhang\emph{,~et~al.}, ``A motion planning and tracking
  framework for autonomous vehicles based on artificial potential field
  elaborated resistance network approach,'' \emph{{IEEE} Transactions on
  Industrial Electronics}, vol.~67, no.~2, pp. 1376--1386, 2020.

\bibitem{malone2017hybrid}
N.~Malone, H.~T. Chiang, K.~Lesser\emph{,~et~al.}, ``Hybrid dynamic moving
  obstacle avoidance using a stochastic reachable set-based potential field,''
  \emph{{IEEE} Transactions on Robotics}, vol.~33, no.~5, pp. 1124--1138, 2017.

\bibitem{lavalle1998rapidly}
S.~LaValle, ``Rapidly-exploring random trees: A new tool for path planning,''
  Computer Science Department, Iowa State University, Tech. Rep. No. 98-11,
  October 1998.

\bibitem{karaman2011sampling}
S.~Karaman and E.~Frazzoli, ``Sampling-based algorithms for optimal motion
  planning,'' \emph{The International Journal of Robotics Research}, vol.~30,
  no.~7, pp. 846--894, 2011.

\bibitem{hu2020efficient}
B.~Hu, Z.~Cao, and M.~Zhou, ``An efficient {RRT}-based framework for planning
  short and smooth wheeled robot motion under kinodynamic constraints,''
  \emph{{IEEE} Transactions on Industrial Electronics}, vol.~68, no.~4, pp.
  3292--3302, 2020.

\bibitem{alvarez2004evolutionary}
A.~Alvarez, A.~Caiti, and R.~Onken, ``Evolutionary path planning for autonomous
  underwater vehicles in a variable ocean,'' \emph{{IEEE} Journal of Oceanic
  Engineering}, vol.~29, no.~2, pp. 418--429, 2004.

\bibitem{niu2020energy}
H.~Niu, Z.~Ji, A.~Savvaris\emph{,~et~al.}, ``Energy efficient path planning for
  unmanned surface vehicle in spatially-temporally variant environment,''
  \emph{Ocean Engineering}, vol. 196, p. 106766, 2020.

\bibitem{ma2018multi}
Y.~Ma, M.~Hu, and X.~Yan, ``Multi-objective path planning for unmanned surface
  vehicle with currents effects,'' \emph{ISA transactions}, vol.~75, pp.
  137--156, 2018.

\bibitem{kufoalor2020autonomous}
D.~K.~M. Kufoalor, T.~A. Johansen, E.~F. Brekke\emph{,~et~al.}, ``Autonomous
  maritime collision avoidance: Field verification of autonomous surface
  vehicle behavior in challenging scenarios,'' \emph{Journal of Field
  Robotics}, vol.~37, no.~3, pp. 387--403, 2020.

\bibitem{huynh2015predictive}
V.~T. Huynh, M.~Dunbabin, and R.~N. Smith, ``Predictive motion planning for
  {AUV}s subject to strong time-varying currents and forecasting
  uncertainties,'' in \emph{2015 IEEE international conference on robotics and
  automation (ICRA)}, 2015, pp. 1144--1151.

\bibitem{fossen2011handbook}
T.~Fossen, \emph{Handbook of Marine Craft Hydrodynamics and Motion
  Control}.\hskip 1em plus 0.5em minus 0.4em\relax Chichester, UK: John Wiley
  and Sons Ltd, 2011.

\bibitem{mueller2015computationally}
M.~Mueller, M.~Hehn, and R.~D'Andrea, ``A computationally efficient motion
  primitive for quadrocopter trajectory generation,'' \emph{{IEEE} Transactions
  on Robotics and Automation}, vol.~31, no.~6, pp. 1294--1310, 2015.

\bibitem{choi2010real}
J.~Choi, R.~Curry, and G.~Elkaim, ``Real-time obstacle-avoiding path planning
  for mobile robots,'' in \emph{Proceedings of the American Institute of
  Aeronautics and Astronautics Conference on Guidance, Navigation, and
  Control}, Toronto, Canada, 2010, p. 8411.

\bibitem{estrada2018forceful}
M.~Estrada, S.~Mintchev, D.~Christensen\emph{,~et~al.}, ``Forceful manipulation
  with micro air vehicles,'' \emph{Science Robotics}, vol.~3, no.~23, 2018.

\bibitem{BlueRobotics2020}
\BIBentryALTinterwordspacing
BlueRobotics. (2021) The {B}lue{R}obotics website. [Online]. Available:
  \url{https://bluerobotics.com}
\BIBentrySTDinterwordspacing

\bibitem{gurobi2019}
\BIBentryALTinterwordspacing
Gurobi. (2020) Gurobi optimizer reference manual. [Online]. Available:
  \url{http://www.gurobi.com}
\BIBentrySTDinterwordspacing

\end{thebibliography}

\end{document}